Title: Chirality of off-center orbital rotation: comparison with lepton spin rotation
Author: Mladen Georgiev (Institute of Solid State Physics, Bulgarian Academy of
  Sciences, 1784 Sofia, Bulgaria)
Comments: 9 pdf pages with 3 figures and 1 appendix
Subj-class: physics

We introduce and attach an orbital chirality to off-center ions in crystals and consider its implications. Its relationship with the spin chirality of leptons and possibly anyons is also discussed. The feature may bring new meaning to the quantum-mechanical vibronic effect.

1. Rationale

Whenever some amount of small-size substitutional impurity is added to an alkali halide host, the impurity ion may go slightly off-center, as far as some tens percent of the interionic separation along one of the crystallographic axes. Among the substitutional ions that displace off-center are the monovalent $Li^+$, $Ag^+$, $Cu^+$, $In^+$, $Ga^+$, $Tl^+$, and aliovalent $Fe^{3+}$ cations as well as the $F^-$ anion in alkali halides, and also the $Mn^{2+}$, $Cu^{2+}$, $Co^{2+}$, $Fe^{2+}$, and $Ni^{2+}$ ions in alkaline earth oxides [1]. The off-center displacements within a centrosymmetric lattice lead to profound changes and a variety of new features in the crystalline host: first and foremost, the off-center effect breaking the inversion symmetry at the normal lattice site, it leads to the occurrence of inversion electric dipoles with the related appearance of paraelectricity [2], to mention a few. On the other hand, the off-center ions perform hindered orbital rotations leading to reorientation around the normal lattice sites which may give rise to magnetic dipoles and thereby to paramagnetism [3]. These orbital rotations are barrier controlled and may be attached a chirality like the one of leptons [4], except for the expectation that the off-center rotation is foremost orbital and not spinning and that it may eventually be reversed through encounter with a barrier.

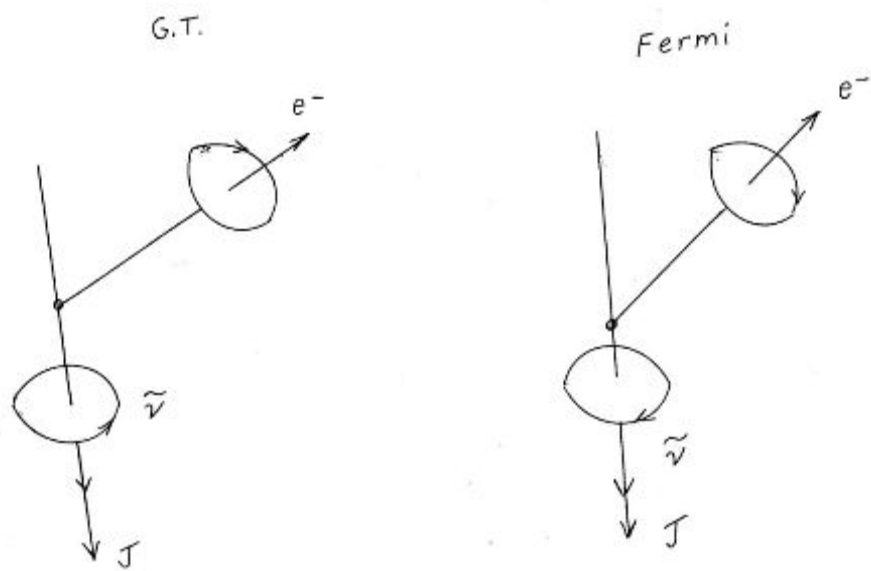

Figure 1(a): Electron $\beta^-$ decay of nucleus showing schematically the momentum of nucleus J, electron $e^-$ and antineutrino $\tilde{\nu}$. Spin rotations (chirality) of leptons are included. Ref. [4].

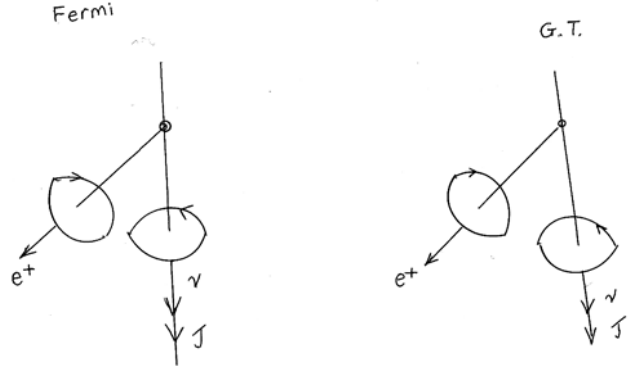

Figure 1(b): Positron $\beta^+$ decay of nucleus showing schematically the momentum of nucleus J, positron e+ and neutrino ν. Spin rotations (chirality) of leptons are included too. Ref. [4].

Nevertheless, the stability of off-center rotation may be evaluated, now and then, making its resemblance to leptons either substantiated or degraded to just a mere formality. This expectation forms the basic scope of the present paper.

## 2. Off-center Hamiltonian

The relevant off-center Hamiltonian accounts in second-quantization terms for the mixing of two nearly-degenerate opposite-parity electron states through coupling to an odd-parity vibrational mode (local mode or phonon mode depending on whether the resulting vibronic polaron is bound or itinerant). The coupling Hamiltonian reads:

$$H_{int} = \Sigma_{\alpha\beta}\, P(G_{\alpha\beta}Q_{\alpha\beta})[a_\alpha^\dagger a_\beta + \text{h.c.}] \qquad (1)$$

where $P(G_{\alpha\beta}Q_{\alpha\beta})$ is an odd-power polynomial of the vibrational coordinates $Q_{\alpha\beta}$ (classic lattice), $G_{\alpha\beta}$ is the coupling constant. (The remaining symbols have been explained at large in the preceding paper [5]). Introducing simplifications and using the premises of small vibrational coordinates, the following form is obtained up to the fourth power in the mode coordinates (lower branch):

$$H_{vib\pm 3D} = -(\hbar^2/2I)(\partial^2/\partial\varphi^2) \pm (K/G)[(D_c-D_b)[(Q_x)^4 + (Q_y)^4 + (Q_z)^4] + D_b] \pm$$

$$E_{JT}[1+(E_{\alpha\beta}/4E_{JT})^2] \qquad (2)$$

K is the stiffness of the (bare) coupled mode, G and D are the electron-mode coupling constants (1st order G, 3rd order D), $E_{JT}$ is the Jahn-Teller energy, $E_{\alpha\beta}$ is the energy gap of the (nearly degenerate) electronic basis states. Introducing spherical coordinates we reduce (1) to:

$$H_{vib\pm 3D} = -(\hbar^2/2I)\Delta_{\theta\varphi} \pm (I\omega^2/G)Q_0^2[(D_c-D_b)[(\cos\varphi\sin\theta)^4 + (\sin\varphi\sin\theta)^4 + (\cos\theta)^4] + D_b] \pm$$

$$E_{JT}[1+(E_{\alpha\beta}/4E_{JT})^2] \qquad (3)$$

where $I = MQ_0^2$ is the inertial moment of a rotating entity of mass M and rotational frequency $\omega_{rot} = \omega\sqrt{[4(D_b-D_c)/G]}Q_0$, $Q_0$ is the off-center radius, viz. the brim radius of the off-center vibronic potential [5]. We note that $D_c$ and $D_b$ are the diagonal matrix elements ($D_c$ principal, $D_b$ lateral) of the coupled constants matrix.

The above forms of Hamiltonians contain both the off-center radius and the reorientational valleys, as well as the rotational barriers in between the reorientational axes (an orientational axis connects the normal lattice site at Q = 0 with the reorientational metastable state at $(\theta,\varphi)_{min}$. Generally, there are eight reorientational sites for the off-center impurity ions in an *fcc* alkali halide lattice.

As it can be seen, the 1st-order-perturbation adiabatic vibronic potential deriving from (1) is a double-branch (±) multidimensional surface. Averaging the angular term in equation (3) (lower branch) over the position angle θ, we obtain the 2-D vibronic potential energy surface (apart from constant terms):

$$E_{vib\pm 3D}(\varphi) \sim \pm (I\omega_{rot}^2)\{½ ¾ ¼ [(\cos\varphi)^4 + (\sin\varphi)^4] + D_b\}$$

$$\sim \pm (I\omega_{rot}^2)\{½ ¾ ¼ ¼ [3 + \cos(4\varphi)] + D_b\} \qquad (4)$$

with a ¾ ¼ ¼ $(I\omega_{rot}^2)$ minimum at $\varphi = 0$ and a ½ ¾ ¼ $(I\omega_{rot}^2)$ maximum at $\varphi = \pi$. The difference gives for the tunneling-barrier height $E_{Brot} = $ ¾ ¼ ¼ $(I\omega_{rot}^2)$ [6].

### 3. Configurational transition probability in hindered orbital rotation

We have just confirmed that the orbital motion along the azimuth coordinate φ is not a smooth one but is rather barrier controlled. The barrier shape is easily seen from equation (4) to be a trigonometric one rather than parabolic, as assumed usually. This makes it necessary to rewrite the math background of the transition probabilities forth and back in the rotation process. In particular, the configurational transition probability along the azimuth coordinate based on the currents across the barrier will be [7]

$$W_{if\,conf}(E_n) = 4\pi^2 |V_{fi}|^2 \sigma_i(E_n)\sigma_f(E_n) \qquad (5)$$

where the matrix element $V_{fi}$ is to be calculated using initial and final state wave functions $u_i$ and $u_f$, respectively, as:

$$V_{fi} = (-\hbar^2/2I) [u_f^* (du_i/d\varphi) - u_i (du_f/d\varphi)^*]|_{\varphi=\varphi_c} \qquad (6)$$

Here $\sigma_i$ and $\sigma_f$ are the corresponding densities (DOS) of the initial and final states.

The rotational states are described by Mathieu's functions. Tthese fall in allowed rotational bands composed of eigenvalues common for the whole off-center rotational sphere (3-D) or ring (2-D). This resulting from the tunneling interaction between the metastable reorientation valleys, it makes the rotational states common too. Under these conditions, the transition probabilities built up of exact eigenstates and eigenvalues would be vanishing which would incapacitate any further analysis based on exact quantities. What prevents this from happening is that the eigenstates and eigenvalues based on Mathieu's quantities are only nearly exact because of the approximations involved in deriving (2) – (4) from (1). On the safe side now, we can revoke Mathieu's functions $Y(\varphi)$ setting $u_i(\varphi) = Y_i(\varphi + ¼\pi)$ and $u_f(\varphi) = Y_f(\varphi - ¼\pi)$, while the DOS are

$$\sigma(E_{a/b,n}) = dn/dE_{a/b,n} = (2I/h^2)(dn/da_n,b_n) \tag{7}$$

since the rotational eigenspectrum is $E_{a/b,n} = (h^2/2I)a_n,b_n$.

Using $Y_{i/f}(\varphi,q)$ we rewrite the saddle-point equation (6) at $\varphi = 0$ to read:

$$V_{if} = (-h^2/2I)\{Y_f(-¼\pi,q)[dY(\varphi.q)/d\varphi]|_{\varphi=¼\pi} - Y(¼\pi,q)[dY_f(\varphi,q)/d\varphi]|_{\varphi=-¼\pi}\} \tag{8}$$

Using expansion in $q$ we calculate finite-valued saddle-point functions at $\varphi = ±¼\pi$. We see that each of the periodic functions $ce_m(z,q)$ and $se_m(z,q)$ is either vanishing or has a vanishing derivative at $\varphi = ¼\pi$. For this reason $ce_m(z,q)$ and $se_m(z,q)$ are not themselves the appropriate rotational eigenstates, though such states may be constructed as linear combinations of basic functions. Such procedures have been employed earlier to compute transition probabilities and reaction rates based on Mathieu's quantities. For greater details the reader is advised to consult references [6-7].

### 4. Probability for changing the electronic state

For transitions under wider crossover gaps $E_{\alpha\beta} \gg h\omega$ (adiabatic transitions) the calculated configuration transition probability $W_{conf}(E_n)$ suffices to making known the actual probability for obtaining a hindered rotation along the off-center ring or sphere. This is the usual situation for rotating entities, especially the ones based on the dynamic Jahn-Teller effect (large gaps). For these rotations the change of electronic state occurs with mere certainty as the rotating species goes from reorientational site to site as the rotation proceeds (not too narrow bands). What does not suffice at $E_{\alpha\beta} \geq h\omega$ is that the electronic state change is increasingly ineffective and may ultimately hold the process. The electronic state change probability is tackled by means of Landau-Zener's method, as described in greater detail elsewhere.

$$W_{el}(E_n) = 2[1 - \exp(-2\pi\gamma(E_n))]/[2 - \exp(-2\pi\gamma(E_n))]$$

$$\gamma(E_n) = (E_{\alpha\beta}^2/2h\omega)[E_R|E_n - E_C|]^{-1/2} \text{ (Landau-Zener's parameter)} \tag{9}$$

which tends to 1 at $\gamma(E_n) \gg 1$. Here $E_R$ is the lattice reorganization energy (the energy involved in creating two reorientational sites), $E_C$ is the crossover energy. We again refer the reader to references for a better account [8-9].

### 5. Total transition probability versus orbital chirality

For conserving space, we assume that the configurational term in the total transition probability

$$W(E_n) = W_{conf}(E_n)W_{el}(E_n) \tag{10}$$

gives the main percentage of efficiency of the tunneling rotation process. In as much as the chirality $\chi_± = ±1$, we define an orbital chirality at energy $E_n$:

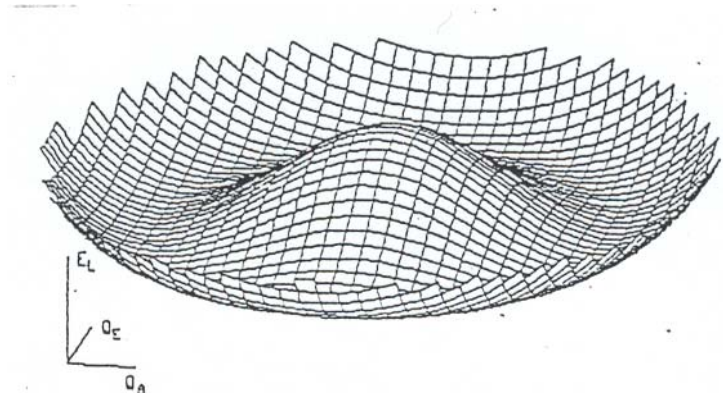

Figure 2: Relief of the vibronic potential showing the off-center path (brim) and the on-center to off-center barrier peaking at the normal lattice site. The relief may be considered representing the off-center chirality, right-handed if the rotation along the brim as looked upon upside down from the barrier top is counter-clock-wise and left-handed vice versa.

$$\chi(E_n)_\pm = \chi_\pm W(E_n) \qquad (11)$$

Here $W(E_n)$ will be regarded renormalized to unity as $W(E_n) / \Sigma_n W(E_n)$. Now the total orbital chirality will also be renormalized: $\Sigma_n \chi(E_n)_\pm = \chi_\pm$ . Here and above the $\pm$ signs should be understood as "either-or" one.

There is no time watch to tell just how many full rotations have materialized counter-clock-wise or clock-wise to produce a $\pm\chi$ chirality. We note that rotations forth and back extending over the same partial number of metastable sites in either direction add up no chirality ($\Delta\chi = 0$). In particular, so do as many transitions forth and back in a two-site situation.

Another problem that is worth mentioning is the pair interactions between rotating species. The latter give rise to magnetic dipoles which are opposite in direction leading to dipole attraction if produced by opposite chiralities or to dipole repulsion if due to parallel chiralities.

One finally wonders whether there is not a link between 2-D off-center impurities or host ions and fractional statistics anyons, as discussed in the literature [10]. If so, new prospects will open for studies of the vibronic effects in solids.

A well-conceived *anyon recipe* envisions three points, among them (i) the respective medium should be an *incompressible fluid*, with (ii) *vertex quasiparticles*, which possess (iii) a *fractional particle number* (leading to fractional statistics) [10]. It seems likely that some if not all of these requirements line up with the off-center species and are to be given a careful consideration. Indeed, we may define a vibronic polaron fluid in lieu of a gas at higher concentrations, while the $Li^+$ vortices may occur in excited $F_A$ state in full-vacancy and/or semi-vacancy configurations like the ones depicted in Figure 4 of reference [5]. The fractional number may materialize too, since a single $Li^+$ ion at $F_A$ in 2-D is distributed amongst four <110> off-center sites and is reduced to ¼ per site. In any event, the anyon connection of vibronic polarons, local or itinerant, is to be addressed more thoroughly in future publications.

Appendix I

Intra- and inter- band transition probabilities

For reasons transparent from the discussion in Reference [11], we consider the following linear combinations of Mathieu's functions $l\{c_m^+(z,q)\} = ce_m(z,q) + ce_{m+1}(z,q)$ and $l\{s_m^+(z,q)\} = se_m(z,q) + se_{m+1}(z,q)$, Equation (8) is redefined in terms of $z = 2\varphi$ to give

$$V_{if,cm} = 2(-h^2/2I_A) \, 2l\{c_m^+(\pi/2,q)\}[(d/dz)lc_m^+(z,q)]|_{z=\pi/2} =$$

$$2(-h^2/2I_A)(d/dz)[l\{c_m^+(z,q)\}]^2|_{z=\pi/2}$$

or, alternatively,

$$V_{if,sm} = 2(-h^2/2I_A) \, 2l\{s_m^+(\pi/2,q)\}[(d/dz)ls_m^+(z,q)]|_{z=\pi/2} =$$

$$2(-h^2/2I_A)(d/dz)[l\{s_m^+(z,q)\}]^2|_{z=\pi/2}$$

The transition matrix element $V_{\{if\}}$ is now finite since for any two consecutive quantum numbers $n = m, m+1$ either a component function ($ce_n(z,q)$ or $se_n(z,q)$) or its derivative is finite at $z = \pi/2$. As examples, we get by even linear combination

$V_{fi,c0} = 2(-h^2/2I_A) \, ce_0(\pi/2,q)[dce_1(\pi/2,q)/dz]$   $V_{fi,c1} = 2(-h^2/2I_A) \, ce_2(\pi/2,q)[dce_1(\pi/2,q)/dz]$

$V_{fi,c2} = 2(-h^2/2I_A) \, ce_2(\pi/2,q)[dce_3(\pi/2,q)/dz]$   $V_{fi,c3} = 2(-h^2/2I_A) \, ce_4(\pi/2,q)[dce_3(\pi/2,q)/dz]$

$V_{fi,c4} = 2(-h^2/2I_A) \, ce_4(\pi/2,q)[dce_5(\pi/2,q)/dz]$,

etc. and also by odd linear combination

$V_{fi,s1} = 2(-h^2/2I_A)\, se_1(\pi/2,q)[dse_2(\pi/2,q)/dz]$   $V_{fi,s2} = 2(-h^2/2I_A)\, se_3(\pi/2,q)[dse_2(\pi/2,q)/dz]$

$V_{fi,s3} = 2(-h^2/2I_A)\, se_3(\pi/2,q)[dse_4(\pi/2,q)/dz]$   $V_{fi,s4} = 2(-h^2/2I_A)\, se_5(\pi/2,q)[dse_4(\pi/2,q)/dz]$

Referring to the preceding discussion, we see that intraband transitions are only nonvanishing if these occur in bands along the upper-sign branch of the adiabatic potential energy surface (APES), such as ones described by the linear combinations

$u(z,q) = \frac{1}{2}[ce_{m-1}(z,q) + ce_m(z,q)] = \frac{1}{2}l\{c_{m-1}^+(z,q)\}$ for m odd

$u(z,q) = \frac{1}{2}[se_{m-1}(z,q) + se_m(z,q)] = \frac{1}{2}l\{s_{m-1}^+(z,q)\}$ for m even

whereas lower-sign branch bands, such as ones described by the linear combinations

$u(z,q) = \frac{1}{2}[ce_{m-1}(z,q) + se_m(z,q)]$ for all m,

do not promote any such transitions.

We first illustrate this statement for the lowest energy bands at "negative q" and "positive q", respectively:

$u_f^*(du_i/dz) - u_i^*(du_f/dz)|_{z=\pi/2} = \frac{1}{2}[ce_0(\pi/2,q) + ce_1(\pi/2,q)]\{[dce_0(\pi/2,q)/dz] + [dce_1(\pi/2,q)/dz]\}$

$= \frac{1}{2}\, ce_0(\pi/2,q)[dce_1(\pi/2,q)/dz]$, q<0,

$u_f^*(du_i/dz) - u_i^*(du_f/dz)|_{z=\pi/2} = \frac{1}{2}\, ce_0(\pi/2,q)[dce_1(\pi/2,q)/dz] - \frac{1}{2}\, se_1(\pi/2,q)[dse_1(\pi/2,q)/dz] = 0$,

q>0, etc. The former generalizes straightforwardly to

$u_f^*(du_i/dz) - u_i^*(du_f/dz)|_{z=\pi/2} =$

$\frac{1}{2}[ce_{m-1}(\pi/2,q) + ce_m(\pi/2,q)]\{[dce_{m-1}(\pi/2,q)/dz] + [dce_m(\pi/2,q)/dz]\} =$

$\frac{1}{4}\, d[ce_{m-1}(z,q) + ce_m(z,q)]^2/dz|_{z=\pi/2}$

for odd m = 1,3,5,... and to

$u_f^*(du_i/dz) - u_i^*(du_f/dz)|_{z=\pi/2} =$

$-\frac{1}{2}[se_{m-1}(\pi/2,q) + se_m(\pi/2,q)]\{[dse_{m-1}(\pi/2,q)/dz] + [dse_m(\pi/2,q)/dz]\} =$

$-\frac{1}{4}\, d[se_{m-1}(z,q) + se_m(z,q)]^2/dz|_{z=\pi/2}$

for even m = 2,4,6,... The relevant transition probabilities are

$W_{if}(E_m) = 4\pi^2 N_m^{-4}\, [2dm/d(a_{m-1} + a_m)]^2\, [ce_{m-1}(\pi/2,q) + ce_m(\pi/2,q)]^2 \times$

$\{[dce_{m-1}(z,q)/dz]|_{z=\pi/2} + [dce_m(z,q)/dz]|_{z=\pi/2}\}^2 =$

$[2dm/d(a_{m-1}+ a_m)]^2 \{d[ce_{m-1}(z,q) + ce_m(z,q)]^2/dz|_{z=\pi/2}\}^2$ (m = 1,3,5,..)

$W_{if}(E_m) = 4\pi^2 N_m^{-4} [2dm/d(b_{m-1}+ b_m)]^2 [se_{m-1}(\pi/2,q) + se_m(\pi/2,q)]^2 \times$

$\{[dse_{m-1}(z,q)/dz]|_{z=\pi/2} + [dse_m(z,q)/dz]|_{z=\pi/2}\}^2 =$

$[2dm/d(b_{m-1}+b_m)]^2 \{d[se_{m-1}(z,q) + se_m(z,q)]^2/dz|_{z=\pi/2}\}^2$ (m = 2,4,6,..)

$W_{if}$ are constructed by linear combinations ½ $1\{c_{m-1}^+(z,q)\}$ and ½ $1\{s_{m-1}^+(z,q)\}$ of normalized eigenfunctions $N_m ce_m(z,q)$ and $N_m se_m(z,q)$ with $N_m = \pi^{-1/2}$, the linear combinations corresponding to the energy eigenvalues $E_m = (h^2/4I_A)(a_{m-1}+a_m)$ for m = 1,3,5,... odd and $E_m = (h^2/4I_A)(b_{m-1}+b_m)$ for m = 2,4,6,... even, respectively.

Strictly speaking, the above linear combination eigenstates corresponding to energy eigenvalues in the middle of the allowed bands, they do not adequately account for the interior of these bands. To improve the description, we make the following proposition: We attach an integer n to number a band where n is odd for $(a_{n-1}, a_n)$ and even for $(b_{n-1}, b_n)$, and let m be a running number $0 \leq m \leq 1$. In so far as the eigenfunctions $ce_m(z,q)$ and $se_m(z,q)$ describing intraband states at noninteger m are not available, we form intraband states by way of linear combinations of band-edge states:

$ce^n_m(z,q) = (1-m)ce_{n-1}(z,q) + m\, ce_n(z,q)$ (n odd)

$se^n_m(z,q) = (1-m)se_{n-1}(z,q) + m\, se_n(z,q)$ (n even)

with intraband eigenvalues

$E^n_m(q) = (1-m)E_{n-1} + mE_n = (h^2/2I)a^n_m(q)$,  $a^n_m(q) = (1-m)a_{n-1}(q) + ma_n(q)$ (n=1,3,5,…odd)

$E^n_m(q) = (1-m)E_{n-1} + mE_n = (h^2/2I)b^n_m(q)$,  $b^n_m(q) = (1-m)b_{n-1}(q) + mb_n(q)$ (n=2,4,6,…even)

respectively. Using the so-constructed intraband states, we redefine the transition probability

$W_{Ln}(E^n_m) = (2\pi)^2 |V^n_m(q)|^2 (dm/dE^n_m)^2$

so as to incorporate

$V^n_m(q) = -(2h^2/I\pi)ce^n_m(z,q)[dce^n_m(z,q)/dz]|_{z=\pi/2} =$

$-(2h^2/I\pi)(1-m)ce_{n-1}(z,q) [mdce_n(z,q)/dz]|_{z=\pi/2}$ (n odd)

$V^n_m(q) = -(2h^2/I\pi)se^n_m(z,q)[dse^n_m(z,q)/dz]|_{z=\pi/2} =$

$-(2h^2/I\pi)(1-m)se_{n-1}(z,q)[mdse_n(z,q)/dz]|_{z=\pi/2}$ (n even)

We get accordingly

$$W_{Ln}(E^n_m) = \begin{cases} 64 \, | (1-m)ce_{n-1}(z,q)[mdce_n(z,q)/dz] |^2_{z=\pi/2} \, (dm/da^n_m)^2 \\ 64 \, | (1-m)se_{n-1}(z,q)[mdse_n(z,q)/dz] |^2_{z=\pi/2} \, (dm/db^n_m)^2 \end{cases}$$

The above probabilities are maximum in the middle of a band at $m = \tfrac{1}{2} W_{Ln}^{max}$ and vanish at the band edges at $m=0$ and $m=1$. To work out an expression feasible for practical calculations the above equation should be normalized to 1. The normalized configuration probabilities are

$$W_{Ln}(E^n_m) = \begin{cases} 64N \, | (1-m)ce_{n-1}(z,q)[mdce_n(z,q)/dz] |^2_{z=\pi/2} \, (dm/da^n_m)^2 \\ 64N \, | (1-m)se_{n-1}(z,q)[mdse_n(z,q)/dz] |^2_{z=\pi/2} \, (dm/db^n_m)^2 \end{cases}$$

where the normalization factor is defined by

$$N^{-1} = 2\sum_{n=1}^{\infty} \int_0^1 W_{Ln}(E^n_m) dm = 128 \sum_{n=1}^{\infty} \int_0^1 dm [m(1-m)]^2 \times$$

$$\begin{cases} |ce_{n-1}(z,q)[dce_n(z,q)/dz]|^2_{z=\pi/2} (dm/da^n_m)^2 \\ |se_{n-1}(z,q)[dse_n(z,q)/dz]|^2_{z=\pi/2} (dm/db^n_m)^2 \end{cases}$$

In cases where Mathieu's functions can be approximated by free-rotor eigenstates $Y_m(\varphi,0) = \pi^{-1/2}\cos(m\varphi)$ we get $V_{fi}(E_n) \sim (h^2/2I\pi)m[-\cos[m(\varphi-\pi/4)] \sin[m(\varphi+\pi/4)] + \cos[m(\varphi+\pi/4)] \times \sin[m(\varphi+\pi/4)]|_{\varphi=0}$ which is equal to $\pm(h^2/2I\pi)m$ for m odd and to 0 for m even. If we set $a_m = m^2$ leading to $\sigma(E_m) = (I/h^2)(1/m)$ we obtain $W_{if}(E_m) = 4\pi^2(h^2/2I\pi)^2 m^2 (I/h^2)^2 (1/m)^2 = 1$ for m odd and $W_{if}(E_m) = 0$ for m even. It implies that the configurational probability of a free rotor is energy-independent, as it should. However if we set $a_m = a_m(q)$ leading to $\sigma(E_n) = (2I/h^2)[dn/da_n(q)]$ we obtain $W_{if}(E_m) = 4\pi^2(h^2n/2I\pi)^2 (2I/h^2)^2 \}(dn/da_n)^2 = 4n^2 (dn/da_n)^2$ which is attributed to quasi-free rotations well above the barrier top.